\documentclass[pre,preprint,superscriptaddress,amsmath,amssymb,floatfix]
{revtex4}
\usepackage{graphics}
\usepackage[dvips]{graphicx}
\begin{document}
\title{Gas-liquid critical parameters of asymmetric models of ionic fluids}
 \author{O.V. Patsahan}
\affiliation{Institute for Condensed Matter Physics of the National
Academy of Sciences of Ukraine, 1 Svientsitskii Str., 79011 Lviv, Ukraine}
 \author{ T.M. Patsahan}
\affiliation{Institute for Condensed Matter Physics of the National
 Academy of Sciences of Ukraine, 1 Svientsitskii Str., 79011 Lviv, Ukraine}
 \date{\today}
\begin{abstract}
The  effects of size and charge asymmetry on the gas-liquid critical
parameters of a primitive model (PM) of ionic fluids  are studied
within  the framework of the statistical field theory based on  the collective
variables method. Recently, this  approach has enabled us to   obtain
the correct trends of the both critical parameters  of the equisize
charge-asymmetric PM without assuming ionic association. In this paper we focus on  the general case
of an asymmetric PM characterized by the two parameters: hard-sphere
diameter-, $\lambda=\sigma_{+}/\sigma_{-}$ and charge,
$z=q_{+}/|q_{-}|$, ratios of the two ionic species. We derive an explicit expression for the chemical potential conjugate to the
order parameter which includes the   effects of correlations up to
the third order. Based on this expression  we consider the three
versions of PM: a monovalent size-asymmetric  PM ($\lambda\neq 1$,
$z=1$), an equisize charge-asymmetric PM ($\lambda=1$, $z\neq 1$)
and a size- and charge-asymmetric PM ($\lambda\neq 1$, $z=2$).
Similar to  simulations, our theory predicts that the critical
temperature and the critical density decrease  with  the increase of size asymmetry.
Regarding the effects of
charge asymmetry, we obtain the correct trend of the critical
temperature with $z$,  while  the trend of the critical density obtained in this approximation is inconsistent 
with simulations, as well as with our previous results
found in the higher-order approximation. We expect that the
consideration of the higher-order correlations will lead to  the
correct  trend  of the critical density with charge asymmetry.
\end{abstract}
\maketitle
\section{Introduction}
It is well-known, that electrostatic forces determine the
properties of various systems: physical as well as chemical or
biological. In particular, the Coulomb interactions are of great
importance when dealing with ionic fluids. Ionic fluids include
molten salts, electrolyte solutions and ionic liquids. In most
cases the Coulomb interaction is the dominant interaction and due
to its long-range character it can substantially affect the
critical properties and the phase behavior of ionic systems. Thus,
the investigations dealing with these issues are of fundamental interest and of practical importance.

Over the last fifteen years, both  phase  diagrams and the critical
behavior of the systems with dominant Coulomb interactions have been
intensively studied by simulations and theoretical methods. A
theoretical model that demonstrates the phase separation driven
exclusively by Coulomb forces is a primitive model (PM). In this
model, the ionic fluid is described as an electroneutral mixture of
charged hard spheres  immersed in a structureless dielectric
continuum. The simplest version of the two-component PM, its
symmetrical version, is called a restricted primitive model (RPM). A
gas-liquid phase transition of the RPM is well established. However, over
the years the figures for the critical parameters have changed
substantially. Now there is a good agreement between the
recent simulations performed by different teams. The  estimations
turn out to be near $T_{c}^{*}\simeq 0.049$, $\rho_{c}^{*}\simeq
0.06-0.08$ when the temperature $T_{c}^{*}$ and the density
$\rho_{c}^{*}$ are in standard dimensionless units
\cite{Hynnien-Panagiotopoulos:08}.

As concerns the theory, several theoretical methods have been
proposed  in which the ion association is  explicitly taken into
account. The main of them are the generalized Debay-H\"uckel (GDH)
theory   and the associated mean spherical approximation
\cite{levinfisher}. These theories are based on the addition of
the chemical association model of Bjerrum \cite{bjerrum} or Ebeling and Grigo \cite{ebeling_grigo1}.
The GDH theory (solvated ion-cluster theory with hard-core term)
yields the following estimations for the critical parameters of
the RPM: $T_{c}^{*}=0.0557$, $\rho_{c}^{*}=0.0261$
\cite{fisher_aqua_banerjee}.

More recently,  the  study of the phase behavior of size- and
charge-asymmetric PMs has been started. The key findings from
simulation studies of asymmetric models are as follows:  the
suitably normalized critical temperatures decrease with size and
charge asymmetry while the critical densities increase with charge
asymmetry but decrease with size asymmetry
\cite{Camp-Patay:99,Romero-Enrique:00,yan-pablo:01,yan-pablo:02,yan-pablo:02:2,panagiot-fisher:02,panagiotopoulos1,cheong-panagiot:03,kim-fisher-panagiotopoulos:05}.
Comparison of simulated critical parameters and theoretical
predictions for asymmetric models has revealed that several
established theories, such as the mean spherical approximation
(MSA)  and the original DH theory are not capable of predicting the
trends observed in simulations \cite{Gonzalez-Tovar,zuckerman:01}. Moreover,
both the original DH theory and the MSA predict no dependence on
charge asymmetry in the equisize case.  The exception are the
theories mentioned above that include the association effects
explicitly
\cite{fisher_aqua_banerjee,Kalyuzhnyj-Holovko-Vlachy:00,artyomov:03}. The
trends found from the GDH theory for the critical parameters of an
equisize ($z$:$1$) charge-asymmetric PM  as a function of charge asymmetry qualitatively agree
with simulation data \cite{fisher_aqua_banerjee}.
As regards the size asymmetry, the extensions of the DH theory  for  monovalent size-asymmetric PMs that  describe the charge-unbalanced "border zones" surrounding  each ion lead to  the trends of the both critical parameters that qualitatively agree  with simulation predictions \cite{zuckerman:01}. However, this is true only for modest size asymmetries. More recently, the  study of the effects of size and charge asymmetry  on the gas-liquid phase separation has been started  within the  field-theoretical description in \cite{Ciach-Gozdz-Stell-07}. It is found that only some of the effects of the size and charge asymmetry are correctly predicted  at the mean-field (MF) level of the theory. For example, the trend of $T_{c}^{*}$ with size asymmetry obtained in this approximation is inconsistent with the predictions of simulations. Summarizing, we
can state that even the  qualitative theoretical understanding of
the issue  is not quite appropriate.

In this paper we focus on the  issue of the effects of charge and
size asymmetry on the gas-liquid critical parameters of PMs. To
this end, we use the statistical field theory based on the method of
collective variables (CVs) (see \cite{Pat-Mryg-CM} and the
references herein).   The approach allows us to derive the exact
functional representation of the grand partition function and
formulate, on this basis, the perturbation theory. Links between
this approach and  the other known theories were established
recently
\cite{patsahan-mryglod-patsahan:06,Cai-Pat-Mryg-CM,patsahan-mryglod:06}.
We also use the method proposed recently
\cite{patsahan-mryglod-patsahan:06}
for the  study of the gas-liquid  phase diagram  of   equisize charge-asymmetric PMs.
The method is based on  determining the chemical potential
conjugate to the  order parameter and allows
one to take into account the effects of higher-order correlations.
Its  application to  an equisize ($z$:$1$) charge-asymmetric PM  enabled us
to obtain the qualitative agreement of the trends of
$T_{c}^{*}(z)$ and $\rho_{c}^{*}(z)$  with simulation findings
\cite{patsahan-mryglod-patsahan:06}. The theory also yields the
best theoretical quantitative estimates for the critical parameters of the RPM \cite{patsahan_ion}. Here we study the general case
of size- and charge-asymmetric PMs.

The layout of the paper is as follows. In Section~2, starting with
the Hamiltonian of a size- and   charge-asymmetric PM we sketch
out the main points of the CVs based theory. We analyze the
Gaussian approximation of the functional of the grand partition
function and determine the CV connected with the order parameter for the gas-liquid
phase separation. In Section~3 we study the gas-liquid critical
parameters of asymmetric PMs  taking into account the correlation
effects of higher order.  We consider the three versions of PM: an
equisize PM with charge asymmetry; a monovalent  PM with size
asymmetry;  a size- and  ($2$:$1$) charge-asymmetric PM. We
conclude in Section~4.

\section{Collective variables based theory for asymmetric PMs}
\subsection{Model}
We consider a classical two-component system consisting of $N_{+}$ hard spheres of diameter $\sigma_{+}$ carrying a charge $q_{+}=zq$ 
and $N_{-}$ hard spheres  of diameter $\sigma_{-}$ carrying a charge $q_{-}=-q$. The ions are immersed in a structureless dielectric medium.
The system is electrically neutral: $\sum_{\alpha=+,-}q_{\alpha}\rho_{\alpha}=0$, $\rho_{\alpha}=N_{\alpha}/V$
is the number density of the $\alpha$th species.

The pair interaction potential is assumed to be of the following form:
\begin{equation}
U_{\alpha\beta}(r)=\phi_{\alpha\beta}^{HS}(r)+\phi_{\alpha\beta}^{C}(r),
\label{2.1a}
\end{equation}
where $\phi_{\alpha\beta}^{HS}(r)$ is the interaction potential between the two  additive hard spheres of diameters $\sigma_{\alpha}$ and $\sigma_{\beta}$. We call the two-component hard sphere system a reference system (RS). Thermodynamic
and structural properties of RS are assumed to be known. $\phi_{\alpha\beta}^{C}(r)$ is the Coulomb potential: $\phi_{\alpha\beta}^{C}(r)=q_{\alpha}q_{\beta}\phi^{C}(r)$, where $\phi^{C}(r)=1/(D r)$, $D$ is the dielectric constant and  hereafter we put $D=1$.
The model is characterized by the parameters of size and charge
asymmetry:
\begin{eqnarray*}
\lambda=\displaystyle\frac{\sigma_{+}}{\sigma_{-}}, \qquad z=\displaystyle\frac{q_{+}}{|q_{-}|}.
\end{eqnarray*}
The fluid is at equilibrium in the grand canonical ensemble. The
grand partition function (GPF) of the model (\ref{2.1a}) can be
written as follows:
\begin{eqnarray*}
\Xi[\nu_{\alpha}]=\sum_{N_{+}\geq 0}\sum_{N_{-}\geq
0}\;\prod_{\alpha=+,-}
\frac{\exp(\nu_{\alpha}N_{\alpha})}{N_{\alpha}!} \int({\rm d}\Gamma)
\exp\left[-\frac{\beta}{2}\sum_{\alpha\beta}\sum_{ij}
U_{\alpha\beta}(r_{ij})\right],
\end{eqnarray*}
where the following notations are used: $\nu_{\alpha}$ is the
dimensionless chemical potential,
$\nu_{\alpha}=\beta\mu_{\alpha}-3\ln\Lambda_{\alpha}$,
$\mu_{\alpha}$ is the chemical potential of the $\alpha$th species,
$\beta$ is the reciprocal temperature, $\Lambda_{\alpha}^{-1}=(2\pi
m_{\alpha}\beta^{-1}/h^{2})^{1/2}$ is the inverse de Broglie thermal
wavelength; $(\rm d\Gamma)$ is the element of configurational space
of the particles.

It is worth  noting that the regularization of the potential
$\phi_{\alpha\beta}^{C}(r)$  inside the hard core is  arbitrary to
some extent.  For example, different regularizations for the Coulomb
potential were considered in \cite{ciach:00:0,caillol_1}. Within the
framework of the Gaussian approximation of GPF the best estimation
for the critical temperature is achieved for the optimized
regularization \cite{anderson_chandler} that leads to the ORPA
(MSA). However, this approximation does not work properly in the
higher orders of the perturbation theory   \cite{caillol_1}. Here we
use the Weeks-Chandler-Andersen  (WCA) regularization scheme
for $\phi_{\alpha\beta}^{C}(r)$ \cite{wcha}:
\begin{equation}
\phi_{\alpha\beta}^{C}(r) = \left\{
                     \begin{array}{ll}
                     q_{\alpha}q_{\beta}/\sigma_{\alpha\beta}, & r< \sigma_{\alpha\beta}\\
                     q_{\alpha}q_{\beta}/r,& r \geq \sigma_{\alpha\beta}.
                     \end{array}
              \right.
\label{WCA}
\end{equation}

\subsection{Functional of the GPF of an asymmetric PM.  The method of collective variables}
Now we use the CVs based theory, developed in \cite{Pat-Mryg-CM} for
a multicomponent continuous system with short- and long-range
interactions in the grand canonical ensemble. As a result,  the
exact functional representation of the GPF for the PM with the
interaction potential (\ref{2.1a}) can be written in  the form:
\begin{equation}
\Xi[\nu_{\alpha}]=\int ({\rm d}\rho)({\rm d}\omega)\exp\left(-{\cal
H}[\nu_{\alpha};\rho_{\alpha},\omega_{\alpha}] \right), \label{2.5}
\end{equation}
where the action ${\cal H}$ reads as
\begin{eqnarray}
{\cal
H}[\nu_{\alpha};\rho_{\alpha},\omega_{\alpha}]=\frac{\beta}{2V}\sum_{\alpha,\beta}\sum_{{\mathbf
k}}\tilde \phi_{\alpha\beta}^{C}(k)\rho_{{\mathbf
k},\alpha}\rho_{-{\mathbf k},\beta}-{\rm
i}\sum_{\alpha}\sum_{{\mathbf k}}\omega_{{\mathbf
k},\alpha}\rho_{{\mathbf k},\alpha}-\ln \Xi_{\rm{HS}}[\bar
\nu_{\alpha}-{\rm i}\omega_{\alpha}]. \label{2.5a}
\end{eqnarray}
Here $\rho_{{\mathbf k},\alpha}=\rho_{{\mathbf k},\alpha}^c-{\rm
i}\rho_{{\mathbf k},\alpha}^s$  is the  CV which describes the value
of the $\mathbf k$-th fluctuation mode of the number density of the
$\alpha$th species, the indices $c$ and $s$ denote real and
imaginary parts of $\rho_{{\mathbf k},\alpha}$.  $\omega_{{\mathbf
k},\alpha}$ is conjugate to the CV $\rho_{{\mathbf k},\alpha}$ and
each of $\rho_{{\mathbf k},\alpha}$ ($\omega_{{\mathbf k},\alpha}$)
takes all the real values from $-\infty$ to $+\infty$. $({\rm
d}\rho)$  and $({\rm d}\omega)$ are  volume elements of the CV phase
space
\begin{displaymath}
({\rm d}\rho)=\prod_{\alpha}{\rm d}\rho_{0,\alpha}{\prod_{\mathbf
k\not=0}}' {\rm d}\rho_{\mathbf k,\alpha}^{c}{\rm d}\rho_{\mathbf
k,\alpha}^{s}, \quad ({\rm d}\omega)=\prod_{\alpha}{\rm
d}\omega_{0,\alpha}{\prod_{\mathbf k\not=0}}' {\rm d}\omega_{\mathbf
k,\alpha}^{c}{\rm d}\omega_{\mathbf k,\alpha}^{s}
\end{displaymath}
and the product over ${\mathbf k}$ is performed in the upper
semi-space ($\rho_{-\mathbf k,\alpha}=\rho_{\mathbf k,\alpha}^{*}$, $\omega_{-\mathbf k,\alpha}=\omega_{\mathbf k,\alpha}^{*}$).

$\tilde \phi_{\alpha\beta}^{C}(k)$ is the Fourier
transform  of the Coulomb potential $\phi_{\alpha\beta}^{C}(r)$. In the case of the WCA regularization (see (\ref{WCA}))  we obtain for $\beta\tilde \phi_{\alpha\beta}^{C}(k)$
\begin{eqnarray}
\beta\tilde\phi_{++}^{C}(k)&=&\frac{4\pi z\sigma_{\pm}^{3}}{T^{*}(1+\delta)}\frac{\sin(x(1+\delta))}{x^{3}}
\label{Coulomb_WCA-1} \\
\beta\tilde\phi_{--}^{C}(k)&=&\frac{4\pi\sigma_{\pm}^{3} }{T^{*}z(1-\delta)}\frac{\sin(x(1-\delta))}{x^{3}}  \label{Coulomb_WCA-2}\\
\beta\tilde\phi_{+-}^{C}(k)&=&-\frac{4\pi\sigma_{\pm}^{3} }{T^{*}}\frac{\sin(x)}{x^{3}},
\label{Coulomb_WCA}
\end{eqnarray}
where  the following notations are introduced:
\begin{equation}
T^{*}=\frac{k_{B}T\sigma_{\pm}}{q^{2}z}
\label{temp}
\end{equation}
is the dimensionless temperature,
\begin{equation}
\delta=\frac{\lambda-1}{\lambda+1}
\label{delta}
\end{equation}
and  $x=k\sigma_{\pm}$,  $\sigma_{\pm}=(\sigma_{+}+\sigma_{-})/2$.

$\Xi_{\rm{HS}}[\bar\nu_{\alpha}-{\rm i}\omega_{\alpha}]$ is the GPF of a
two-component hard sphere system  with the renormalized
chemical potential
\begin{eqnarray*}
\bar \nu_{\alpha}=\nu_{\alpha}+\frac{\beta}{2V}\sum_{{\mathbf
k}}\tilde\phi_{\alpha\alpha}^{C}(k)
\end{eqnarray*}
in the presence of the local field $-{\rm i}\omega_{\alpha}(r)$.

In order to formulate the perturbation theory we present the CVs in
the following form:
\[
\rho_{{\mathbf k},\alpha}=\bar\rho_{\alpha}\delta_{{\mathbf
k}}+\delta\rho_{{\mathbf k},\alpha}, \quad \omega_{{\mathbf
k},\alpha}=\bar\omega _{\alpha}\delta_{{\mathbf
k}}+\delta\omega_{{\mathbf k},\alpha},
\]
where the mean-field (MF) values $\bar\rho_{\alpha}$ and $\bar\omega _{\alpha}$ are the solutions of the saddle point equations.

Then we  present  $\ln\Xi_{\rm{HS}}[\bar\nu_{\alpha}-{\rm
i}\omega_{\alpha}]$ in (\ref{2.5a}) in the form of the cumulant
expansion
\begin{eqnarray}
\ln\Xi_{\rm{HS}}[\ldots]&=&\sum_{n\geq 0}\frac{(-{\rm
i})^{n}}{n!}\sum_{\alpha_{1},\ldots,\alpha_{n}}
\sum_{{\mathbf{k}}_{1},\ldots,{\mathbf{k}}_{n}}
{\mathfrak{M}}_{\alpha_{1}\ldots\alpha_{n}}(\bar\nu_{\alpha}-{\rm
i}\bar\omega_{\alpha}; k_{1},\ldots,k_{n})
\delta\omega_{{\bf{k}}_{1},\alpha_{1}}\ldots\delta\omega_{{\bf{k}}_{n},\alpha_{n}}
\nonumber\\
&&\times \delta_{{\bf{k}}_{1}+\ldots +{\bf{k}}_{n}}, \label{2.11}
\end{eqnarray}
with  ${\mathfrak{M}}_{\alpha_{1}\ldots\alpha_{n}}(\bar\nu_{\alpha}-{\rm
i}\bar\omega_{\alpha}; k_{1},\ldots,k_{n})$ being the $n$th cumulant defined by
\begin{equation}
{\mathfrak{M}}_{\alpha_{1}\ldots\alpha_{n}}(\bar\nu_{\alpha}-{\rm
i}\bar\omega_{\alpha};k_{1},\ldots,k_{n})=\left.
\frac{\partial^{n}\ln \Xi_{\rm{HS}}[\ldots]}{
\partial\delta\omega_{{\bf{k}}_{1},\alpha_{1}}\ldots\partial\delta\omega_{{\bf{k}}_{n},\alpha_{n}}}\right\vert_{\delta\omega_{{\bf{k}}_{i},\alpha_{i}}=0}.
\label{2.12}
\end{equation}
The $n$th cumulant ${\mathfrak{M}}_{\alpha_{1}\ldots\alpha_{n}}$
coincides with the Fourier transform of the $n$-partical connected correlation function of the RS \cite{Pat-Mryg-CM}.
$\delta_{{\bf{k}}_{1}+\ldots+{\bf{k}}_{n}}$ is the Kronecker symbol.
The $n$th cumulant  depends on both the wave vectors $\bf
k_{i}$ and the partial chemical potentials $\bar
\nu_{\alpha}-{\rm
i}\bar\omega_{\alpha}$.

Using (\ref{2.11})-(\ref{2.12}) we can rewrite
(\ref{2.5})-(\ref{2.5a}) as follows
\begin{eqnarray}
\Xi[\nu_{\alpha}]&=&\Xi_{\rm{MF}}[\bar\nu_{\alpha}-{\rm
i}\bar\omega_{\alpha}]\int({\mathrm{d}}\delta\rho)
\exp\Big\{-\frac{\beta}{2V}\sum_{\alpha,\beta}\sum_{\bf
k}\tilde\phi_{\alpha\beta}^{C}(k)\delta\rho_{{\bf
k},\alpha}\delta\rho_{-{\bf k},\beta}\nonumber \\
&& 
+{\rm
i}\sum_{\alpha}\sum_{{\mathbf k}}\delta\omega_{{\mathbf
k},\alpha}\delta\rho_{{\mathbf
k},\alpha}+\sum_{n\geq 2}\frac{(-{\rm
i})^{n}}{n!}\sum_{\alpha_{1},\ldots,\alpha_{n}}
\sum_{{\mathbf{k}}_{1},\ldots,{\mathbf{k}}_{n}}
{\mathfrak{M}}_{\alpha_{1}\ldots\alpha_{n}}(\bar\nu_{\alpha}-{\rm
i}\bar\omega_{\alpha}; k_{1},\ldots,k_{n})
\nonumber\\
&&\times \delta\omega_{{\bf{k}}_{1},\alpha_{1}}\ldots\delta\omega_{{\bf{k}}_{n},\alpha_{n}}
\delta_{{\bf{k}}_{1}+\ldots +{\bf{k}}_{n}}\Big\},
\label{dA.14}
\end{eqnarray}
where $\Xi_{\rm{MF}}$ is the GPF of the model in the MF approximation.

\paragraph{Gaussian approximation}
Now we consider the Gaussian approximation of $\Xi[\nu_{\alpha}]$
setting  ${\mathfrak{M}}_{\alpha_{1}\ldots\alpha_{n}}\equiv 0$ for
$n\geq 3$. Then, after integration in (\ref{dA.14})
over $\delta\omega_{{\bf{k}},\alpha}$ we obtain
\begin{eqnarray}
\Xi_{{\text G}}[\nu_{\alpha}]=\Xi_{\rm{MF}}[\bar\nu_{\alpha}-{\rm
i}\bar\omega_{\alpha}]\;\Xi'\int(\mathrm{d}\delta\rho)
\exp\Big\{-\frac{1}{2}\sum_{\alpha,\beta}\sum_{\bf
k}\tilde{\cal C}_{\alpha\beta}(k)\delta\rho_{{\bf k},\alpha}\delta\rho_{-{\bf
k},\beta}\Big\},
\label{Ksi-G}
\end{eqnarray}
where $\tilde{\cal C}_{\alpha\beta}(k)$ is the Fourier transform of the two-particle direct correlation function in the Gaussian approximation
\begin{equation}
\tilde{\cal
C}_{\alpha\beta}(k)=\frac{\beta}{V}\tilde\phi_{\alpha\beta}^{C}(k)+
\frac{1}{\sqrt{N_{\alpha}N_{\beta}}}\tilde{\cal
C}_{\alpha\beta}^{\text HS}(k). \label{C-alpha-beta}
\end{equation}
$\tilde{\cal C}_{\alpha\beta}^{\text HS}(k)$ is the
Fourier transform of the direct correlation function of a
two-component hard-sphere system. It is connected with
${\mathfrak{M}}_{\alpha\beta}(k)$ by the relation
\begin{equation}
\tilde {\cal C}_{2}^{{\text
HS}}(k){\mathfrak{M}}_{2}(k)=\underline{1}, \label{O-Z}
\end{equation}
where $\tilde {\cal C}_{2}^{{\text HS}}(k)$ denotes the matrix of
elements  $\tilde{\cal C}_{\alpha\beta}^{\text
HS}(k)/\sqrt{N_{\alpha}N_{\beta}}$ and ${\mathfrak{M}}_{2}$ the matrix of elements ${\mathfrak{M}}_{\alpha\beta}(k)$. $\underline{1}$ is
the unit matrix. It should be noted that  $\tilde{\cal C}_{\alpha\beta}(k)$ is connected to the ordinary direct correlation function $\tilde c_{\alpha\beta}(k)$  by \cite{hansen_mcdonald}
\begin{displaymath}
\tilde{\cal C}_{\alpha\beta}(k)=\frac{\delta_{\alpha\beta}}{<\rho_{\alpha}>}-\tilde c_{\alpha\beta}(k),
\end{displaymath}
where  $\rho_{\alpha}=<N_{\alpha}>/V$.

In order to determine the  CV connected with the order parameter we follow the ideas of \cite{Ciach-Gozdz-Stell-07,Patsahan-Patsahan} and introduce independent collective excitations  by means of the orthogonal transformation
\begin{eqnarray}
\delta\rho_{{\mathbf k},+}&=&A(k)\xi_{{\mathbf k},1}+C(k)\xi_{{\mathbf k},2} \nonumber\\
\delta\rho_{{\mathbf k},-}&=&B(k)\xi_{{\mathbf k},1}+D(k)\xi_{{\mathbf k},2}.
\label{xi-i}
\end{eqnarray}
The explicit form of coefficients $A(k)$, $B(k)$, $C(k)$ and $D(k)$ are given in Appendix~A. As a result, (\ref{Ksi-G}) is rewritten as
\begin{eqnarray}
\Xi_{{\text G}}[\nu_{\alpha}]=\Xi_{\rm{MF}}[\bar\nu_{\alpha}-{\rm
i}\bar\omega_{\alpha}]\;\Xi'\int(\mathrm{d}\xi)
\exp\Big\{-\frac{1}{2}\sum_{\alpha=1,2}\sum_{\bf
k}\tilde\varepsilon_{\alpha}(k)\xi_{{\bf k},\alpha}\xi_{-{\bf
k},\alpha}\Big\},
\label{Ksi-xi}
\end{eqnarray}
where eigenvalues $\tilde\varepsilon_{1}(k)$ and
$\tilde\varepsilon_{2}(k)$ are found to be
\begin{equation}
\tilde\varepsilon_{1,2}(k)=\frac{1}{2}\left(\tilde{\cal C}_{++}(k)+\tilde{\cal C}_{--}(k)\pm  \left[(\tilde{\cal C}_{++}(k)-\tilde{\cal C}_{--}(k))^{2}+4\tilde{\cal C}_{+-}^{2}(k)\right]^{1/2}\right).
\label{epsilon-i}
\end{equation}
Here we are interested in the gas-liquid critical point. Thus, we
are now in a position to study equations (\ref{xi-i}) and
(\ref{epsilon-i}) in the long-wavelength limit. In this case
equations (\ref{xi-i}) have the form
\begin{eqnarray}
\delta\rho_{0,+}&=&\frac{z}{\sqrt{1+z^2}}\xi_{0,1}+\frac{1}{\sqrt{1+z^{2}}}\xi_{0,2}, \nonumber\\
\delta\rho_{0,-}&=&-\frac{1}{\sqrt{1+z^2}}\xi_{0,1}+\frac{z}{\sqrt{1+z^{2}}}\xi_{0,2}
\label{xi-0}
\end{eqnarray}
which in turn leads to the relations
\begin{eqnarray}
\xi_{0,1}&=&\frac{1}{\sqrt{1+z^2}}\left(z\delta\rho_{0,+}-\delta\rho_{0,-}\right) , \nonumber\\
\xi_{0,2}&=&\frac{1}{\sqrt{1+z^2}}\left(\delta\rho_{0,+}+z\delta\rho_{0,-}\right).
\label{rho-0}
\end{eqnarray}
Introducing CVs $\rho_{0,N}=\delta\rho_{0,+}+\delta\rho_{0,-}$ and
$\rho_{0,Q}=z\delta\rho_{0,+}-\delta\rho_{0,-}$ that describe
long-wavelength fluctuations of the total number density and charge
density, respectively, we can
rewrite (\ref{rho-0}) in the form
\begin{eqnarray}
\xi_{0,1}&=&\frac{1}{\sqrt{1+z^2}}\rho_{0,Q}, \nonumber\\
\xi_{0,2}&=&\frac{1}{\sqrt{1+z^2}}\left(\frac{1+z^{2}}{1+z}\rho_{0,N}+\frac{1-z}{1+z}\rho_{0,Q}\right).
\label{rho1-0}
\end{eqnarray}
As is seen, CV $\xi_{0,1}$  describes
fluctuations of the charge density. In the general case $z\neq 1$,   $\xi_{0,2}$
is a linear combination of CVs $\rho_{0,N}$ and $\rho_{0,Q}$ with the z-dependent coefficients. At  $z= 1$, CV $\xi_{0,2}$ describes solely fluctuations of the total number density. Thus, we suggest that CV $\xi_{0,2}$ is connected with the order
parameter of the gas-liquid critical point.

At $k=0$ one finds that
\begin{equation}
\tilde\varepsilon_{1}(k=0)=\infty,
\label{epsilon1-0}
\end{equation}
\begin{eqnarray}
\tilde\varepsilon_{2}(k=0)=\frac{1+z}{1+z^2}\left(-\frac{4\pi\rho^{*}z\delta^{2}}{3T^{*}(1+z)}+\tilde
c_{++}^{{\text HS}}(0)+2\sqrt{z} \tilde c_{+-}^{{\text
HS}}(0)+z\tilde c_{--}^{{\text HS}}(0)\right), \label{epsilon2-0}
\end{eqnarray}
where $T^{*}$ and $\delta$ are given by (\ref{temp})-(\ref{delta})
and $\rho^{*}=\rho\sigma_{\pm}^{3}$ is a reduced total number
density.

Equation (\ref{epsilon1-0}) leads to
\begin{equation}
\tilde G_{QQ}(k=0)=0,
\label{G-QQ}
\end{equation}
where $\tilde G_{QQ}(k=0)$ is the Fourier transformation of the charge-charge connected correlation function;  equation (\ref{G-QQ}) reflects the fact that the first moment  Stillinger-Lovett rule is satisfied in the Gaussian approximation.

At $\delta=0$, $\tilde\varepsilon_{2}(k=0)$ reduces to the form
\begin{equation*}
\tilde\varepsilon_{2}(\delta=0;k=0)=\frac{(1+z)^{2}}{1+z^2}\frac{1}{S_{2}(0)},
\end{equation*}
where $S_{2}(0)$ is the two-particle structure factor of a one-component
hard-sphere system at $k=0$. In the Percus-Yevick (PY) approximation
\cite{hansen_mcdonald}
\begin{equation}
S_{2}(0)=\frac{(1-\eta)^{4}}{(1+2\eta)^{2}}, \label{S-2}
\end{equation}
where $\eta=\displaystyle{\frac{\pi }{6}}\rho\sigma^{3}$ is the
packing fraction. It is worth noting  that
$\tilde\varepsilon_{2}(\delta=0;k=0)$ takes only positive values. It suggests that in the size-symmetric case no  phase separation between two uniform phases can be found at the Gaussian level of the description.

Equation $\tilde\varepsilon_{2}(\delta\neq 0;k=0)=0$ leads to the
gas-liquid spinodal curve in the Gaussian approximation
\begin{equation}
T^{*}_{s}=\frac{4\pi\rho^{*}\delta^{2}}{3(1+z)}\left(\tilde
c_{++}^{{\text HS}}(0)/z+2 \tilde c_{+-}^{{\text HS}}(0)/
\sqrt{z}+\tilde c_{--}^{{\text HS}}(0) \right)^{-1}. \label{T-s}
\end{equation}
Equations (\ref{xi-0})-(\ref{T-s}) are analogous to those obtained
in \cite{Ciach-Gozdz-Stell-07} but for another type of the
regularization of the Coulomb potential inside the hard core. The
trends of the critical parameters calculated from the maximum point
of  spinodal (\ref{T-s}) are consistent with the corresponding
trends  found in \cite{Ciach-Gozdz-Stell-07}: at the fixed $z$ the
critical temperature $T_{c}^{*}$ is a convex down  function of
$\delta$ while the critical density $\rho^{*}(\delta)$ is  a convex
up in $\delta$;  both $T^{*}_{c}$ and $\rho^{*}_{c}$ increase at a
given $\delta>0$ and decrease at a given $\delta<0$ when   $z$
increases. Therefore, only some of the trends are correctly predicted within the framework of this approximation. In order to properly describe  the effects  of
size and charge asymmetry on the critical parameters one should take
into account  the terms of the higher-order than the second order
in the functional Hamiltonian (\ref{dA.14}). We consider this  task
below.

\section{Critical parameters of  asymmetric PMs: Beyond the Gaussian approximation}

In order to study  the gas-liquid critical points  of  asymmetric
PMs we use the method proposed in
\cite{patsahan-mryglod-patsahan:06}.
First we pass from the initial chemical potentials $\nu_{+}$ and
$\nu_{-}$ to their linear combinations
\begin{equation}
\nu_{1}=\frac{z\nu_{+}-\nu_{-}}{\sqrt{1+z^{2}}}, \qquad
\nu_{2}=\frac{\nu_{+}+z\nu_{-}}{\sqrt{1+z^{2}}}.
\label{2.8a}
\end{equation}
Chemical potentials  $\nu_{1}$ and $\nu_{2}$  are conjugate
to CVs $\xi_{0,1}$  and $\xi_{0,2}$, respectively. 
Since we suggest that CV $\xi_{0,2}$ is connected with the order parameter, $\nu_{2}$  appears to be  of special interest in our study.

Following the ideas of \cite{patsahan-mryglod-patsahan:06} we start with
the logarithm of GPF  in the Gaussian approximation (\ref{Omega-G})
\begin{equation}
\ln\Xi_{G}[\nu_{\alpha}]=\ln\Xi_{\rm{HS}}[\bar\nu_{\alpha}]-\frac{1}{2}\sum_{\mathbf
k}\ln{\det}\,\left[\underline{1}+\Phi_{C}{\mathfrak{M}}_{2}\right],
\label{Omega-G}
\end{equation}
where $\Phi_{C}$ and ${\mathfrak{M}}_{2}$ are  matrices of elements
$\beta\tilde\phi_{\alpha\beta}^{C}(k)$ and
${\mathfrak{M}}_{\alpha\beta}(\bar\nu_{\alpha}-{\rm i}\bar\omega_{\alpha};
k)$, respectively.

We  approximate  cumulants ${\mathfrak{M}}_{\alpha\beta}(k)$   by their values in the  long-wavelength limit putting  ${\mathfrak{M}}_{\alpha\beta}(k)={\mathfrak{M}}_{\alpha\beta}(k=0)={\mathfrak{M}}_{\alpha\beta}$.
If it is remembered that $\ln\Xi_{\rm{HS}}$ and ${\mathfrak{M}}_{\alpha_{1}\alpha_{2}\ldots\alpha_{n}}$ are functions of the full  chemical potentials   we can present $\nu_{1}$ and $\nu_{2}$  as
\[
\nu_{1}=\nu_{1}^{0}+\lambda^{0}\Delta\nu_{1}, \qquad
\nu_{2}=\nu_{2}^{0}+\lambda^{0}\Delta\nu_{2}, \qquad
\]
with $\nu_{1}^{0}$ and $\nu_{2}^{0}$ being the MF values of
$\nu_{1}$ and $\nu_{2}$, respectively and $\Delta\nu_{1}$
and $\Delta\nu_{2}$  being the solutions of the equations
\begin{eqnarray}
\frac{\partial\ln
\Xi_{G}(\nu_{1},\nu_{2})}{\partial\Delta\nu_{1}}&=0 ,
\label{a3.17}
\\
\frac{\partial\ln
\Xi_{G}(\nu_{1},\nu_{2})}{\partial\Delta\nu_{2}}&=&\lambda^{0}\left(\langle
N_{+}\rangle_{HS}+z\langle N_{-}\rangle_{HS} \right) .
\label{b3.17}
\end{eqnarray}

We  self-consistently solve  equations (\ref{a3.17})-(\ref{b3.17})
for the relevant chemical potential $\Delta\nu_{2}$ by means of
successive approximations keeping terms of a certain order in
parameter $\lambda^{0}$ \cite{patsahan-mryglod-patsahan:06}. To
this end, we expand (\ref{Omega-G}) in powers of $\Delta\nu_{1}$
and $\Delta\nu_{2}$
\begin{equation}
\ln\Xi_{G}(\nu_{1},\nu_{2})=\sum_{n\geq 0}\sum_{i_{n}\geq 0}^{n}C_{n}^{i_{n}} \frac{{\mathcal
M}_{n}^{(i_{n})}(\nu_{11}^{0},\nu_{2}^{0})}{n!}\Delta\nu_{1}^{n-i_{n}}\Delta\nu_{2}^{i_{n}},
\label{a3.16}
\end{equation}
where
\[
\left. {\cal{M}}_{n}^{(i_{n})}(\nu_{1}^{0},\nu_{2}^{0})= \frac{\partial^{n}\ln\Xi_{G}(\nu_{1},\nu_{2})}{\partial\Delta\nu_{1}^{n-i_{n}}\partial\Delta\nu_{2}^{i_{n}}}\right\vert_{\Delta\nu_{1}=0,\Delta\nu_{2}=0}.
\]
The first nontrivial approximation  corresponding to
$\Delta\nu_{1}=0$ yields the following expression for
$\Delta\nu_{2}$
\begin{eqnarray}
\Delta\nu_{2}&=&\frac{\sqrt{1+z^{2}}}{2V\left[{\mathfrak{M}}_{++}+2z{\mathfrak{M}}_{+-}
+z^{2}{\mathfrak{M}}_{--}\right]}\sum_{{\mathbf k}}\frac{1}{{\text{det}}\,[\underline{1}+\Phi_{C}{\mathfrak{M}}_{2}]}\left(\beta\tilde\phi_{++}(k){\cal S}_{1}
\right.
\nonumber
\\
&&
\left.
+\beta\tilde\phi_{--}(k){\cal S}_{2}+2\beta\tilde\phi_{+-}(k){\cal S}_{3}\right),
\label{delta-nu2}
\end{eqnarray}
where
\begin{equation}
{\cal S}_{1}={\mathfrak{M}}_{+++}+z{\mathfrak{M}}_{++-}, \qquad {\cal
S}_{2}={\mathfrak{M}}_{+--}+z{\mathfrak{M}}_{---},\qquad {\cal
S}_{3}={\mathfrak{M}}_{++-}+z{\mathfrak{M}}_{+--}.
\label{S-3}
\end{equation}
Apart from   ${\mathfrak{M}}_{\alpha_{1}\alpha_{2}}$, formulas  (\ref{delta-nu2})-(\ref{S-3}) include  the third order cumulants ${\mathfrak{M}}_{\alpha_{1}\alpha_{2}\alpha_{3}}$ or equivalently the third order connected correlation functions of the RS.

Finally, we can write  the full chemical potential  $\nu_{2}$ in the form
\begin{equation}
\nu_{2}=\nu_{2}^{HS}+\nu_{2}^{S}+\Delta\nu_{2},
\label{chem-pot-full}
\end{equation}
where
\begin{equation}
\nu_{2}^{HS}=\frac{\nu_{+}^{HS}+z\nu_{-}^{HS}}{\sqrt{1+z^{2}}}
\label{nu_HS}
\end{equation}
with $\nu_{+}^{HS}$ ($\nu_{-}^{HS}$) being the hard-sphere chemical potential  of the $\alpha$th species and $\nu_{2}^{S}$ being the combination of the self-energy parts of  chemical potentials $\nu_{+}$ and $\nu_{-}$
\begin{equation}
\nu_{2}^{S}=-\frac{1}{2V\sqrt{1+z^{2}}}\sum_{{\mathbf
k}}\left( \beta\tilde\phi_{++}^{C}(k)+z\beta\tilde\phi_{--}^{C}(k)\right).
\label{nu_s}
\end{equation}
Now some comments are in order:
\begin{itemize}
\item Here we consider the two-component hard-sphere system  as a
RS. In this case the analytical expressions for second order
cumulants ${\mathfrak{M}}_{\alpha_{1}\alpha_{2}}$   can be
obtained in the PY approximation using the
Lebowitz' solution \cite{leb1,lebrow}. The corresponding formulas
for ${\mathfrak{M}}_{\alpha_{1}\alpha_{2}}(k=0)$  are given in
Appendix~B. \item In order to derive the expressions for the third
order cumulants one can use the recurrent relation
\begin{equation}
{\mathfrak{M}}_{\alpha_{1}\alpha_{2}\ldots\alpha_{n}}={\mathfrak{M}}_{\alpha_{1}\alpha_{2}\ldots\alpha_{n}}(0,\ldots)=\frac{\partial{\mathfrak{M}}_{\alpha_{1}\alpha_{2}\ldots\alpha_{n-1}}(0,\ldots)}{\partial\nu^{0}_{\alpha_{n}}},
\label{nth-cumulant}
\end{equation}
where $\nu^{0}_{\alpha_{i}}$ is the MF value of chemical potential
$\nu_{\alpha_{i}}$ which due to the electroneutrality condition
coincides   with  the hard-sphere  chemical potential  of the
$\alpha_{i}$th species.
\end{itemize}

Formulas (\ref{delta-nu2})-(\ref{nu_s}) will be used  for the
study of the gas-liquid phase equilibria  in asymmetric PMs. Below
we consider some particular cases.

\subsection{Monovalent PMs with size asymmetry}

First we consider a monovalent PM with size asymmetry corresponding
to $z=1$ and $\lambda\neq 1$. Because of symmetry with respect to
the exchange of $+$ and $-$ ions, only $\lambda<1$ (or $\lambda>1$)
need be considered in this case.

We put $z=1$ in (\ref{delta-nu2})-(\ref{nu_s})  and consider the PY
approximation for the thermodynamic and structural functions of the
two-component hard sphere system \cite{leb1,lebrow}.  For cumulants
${\mathfrak{M}}_{\alpha_{1}\alpha_{2}}$ and
${\mathfrak{M}}_{\alpha_{1}\alpha_{2}\alpha_{3}}$ we use formulas
(\ref{c11.a})-(\ref{S112}) from Appendix~B. The Fourier transforms
of the interaction potentials  are given by
(\ref{Coulomb_WCA-1})-(\ref{Coulomb_WCA}).

The explicit expressions
for $\nu_{2}^{\text HS}$ and $\nu_{2}^{\text S}$ are obtained
using the results of Ref. \cite{lebrow} supplemented by the
electroneutrality condition. They are given in Appendix~C.

Based on the expressions  (\ref{delta-nu2})-(\ref{nu_s}) (at $z=1$) supplemented by the Maxwell construction  we calculate the
coexistence curves and the corresponding critical parameters for
different values of $\lambda$.  Estimates of the critical point
values of $T_{c}^{*}$ and $\rho_{c}^{*}$ are given by their values
for which the maxima and minima of the van der Waals loops
coalesce. The estimated values of the  critical parameters
are presented in table~1.
\begin{table}[htbp]
\caption{Critical parameters $T_{c}^{*}=k_{B}T\sigma_{\pm}/q^{2}$ and
$\rho_{c}^{*}=\rho_{c}\sigma_{\pm}^{3}$ of the monovalent PM for different
values of $\lambda$} \vspace{3mm}
\begin{tabular}{ccc}
\hline \hline
\hspace{10mm}  $\lambda$\hspace{10mm} &
\hspace{10mm} $T_{c}^{*}$ \hspace{10mm}&
\hspace{10mm}$10^{2}\rho_{c}^{*}$\hspace{10mm} \\
\hline
$1.0$  & $0.0848$ & $0.907$\\
$0.75$ &$0.0831$ &$0.816$ \\
$0.5$ &  $0.0786$  &$0.637$ \\
$0.25$ & $0.0709$& $0.433$\\
$0.1$& $0.0586$ & $0.195$\\
\hline
\hline
\end{tabular}
\end{table}

\begin{figure}[h]
\centering
\includegraphics[height=6.5cm]{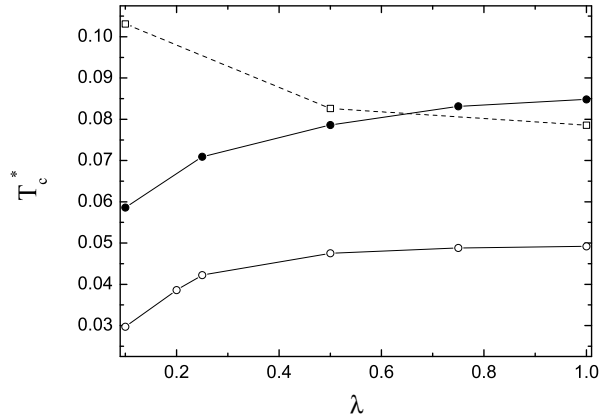}
\caption{Critical temperature $T_{c}^{*}$ of the monovalent PM as a
function of size asymmetry. Open circles correspond to the results of
simulations \cite{yan-pablo:02}; open squares are
MSA results via the energy route \cite{Gonzalez-Tovar} and solid circles correspond to the results of the CV based theory.} 
\end{figure}
Figures 1 and 2 demonstrate the effects of size asymmetry on the
critical parameters of the monovalent PM. In Fig.~1 the critical
temperature $T_{c}^{*}$ depending on  $\lambda$ is shown  by  the
solid circles for $\lambda$ ranging from  $0.1$ to $1$. As is seen,
a qualitative agreement with the simulation data  shown by the open
circles is obtained. The results obtained in  the MSA are shown by
the open squares. In Fig~2 the dependence of the critical density
$\rho_{c}^{*}$ on $\lambda$ is shown. Similar to the computer
simulation findings our results indicate a decrease of the critical
density with the increase of $\lambda$ but the figures obtained in
this approximation turn out to be too small.
\begin{figure}[h]
\centering
\includegraphics[height=6.5cm]{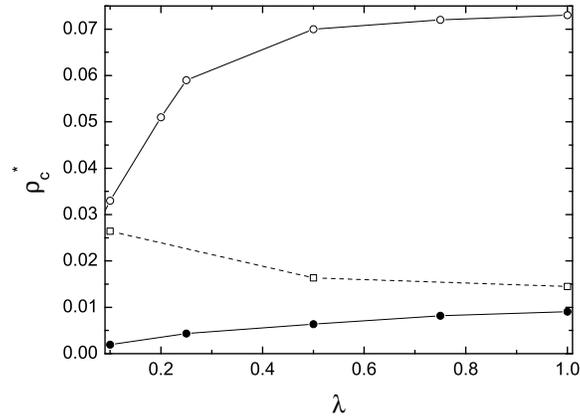}
\caption{Critical density $\rho_{c}^{*}$ of the monovalent PM as a
function of size asymmetry. The meaning of
the symbols is the same as in Fig.~1.} 
\end{figure}

It should be noted that  the RPM limit turns out to be a special
case.  When  $z=\lambda=1$ expression (\ref{delta-nu2})
reduces to the form that corresponds to the random phase
approximation (RPA) \cite{patsahan-mryglod-patsahan:06}.

\subsection{Equisize PMs with charge asymmetry}
Another particular case is an equisize PM with charge asymmetry
corresponding to $\lambda=1$ and $z\neq 1$. At $\lambda=1$
($\delta=0$) the expression
\[
\beta\tilde\phi_{++}(k){\cal S}_{1}
+\beta\tilde\phi_{--}(k){\cal S}_{2}+2\beta\tilde\phi_{+-}(k){\cal S}_{3}
\]
entering (\ref{delta-nu2}) reduces to the form (see Eqs.~(\ref{Coulomb_WCA-1})-(\ref{Coulomb_WCA}),
(\ref{delta-nu2})-(\ref{S-3}) and the formulas in Appendix~B):
\begin{equation}
-\frac{\sin(x)}{T^{*}x^{3}}\left( (1-z)^{2}-(1+z^{2})S_{2}(0)\right),
\label{reduce_charge}
\end{equation}
where  the dimensionless temperature $T^{*}$ is given by Eq.~(\ref{temp}) under condition $\sigma_{\pm}=\sigma$.  $S_{2}(0)$ is
given by (\ref{S-2}).

As a result, Eqs.~(\ref{delta-nu2})-(\ref{nu_s}) read as
\begin{equation}
\Delta\nu_{2}=\frac{1+z}{\sqrt{1+z^{2}}}\frac{i_{1}}{\pi}\left(1-\frac{(1-z^{3})(1-z)}{z(1-z)^{2}+(1+z^{2})^{2}S_{2}(0)} \right),
\label{deltanu-lambda-1}
\end{equation}
\begin{equation}
\nu_{2}^{HS}=\frac{(1+z)}{\sqrt{1+z^{2}}}\nu^{HS}, \qquad
\nu_{2}^{S}=-\frac{1+z}{\sqrt{1+z^{2}}}\frac{1}{2T^{*}},
\label{nu-lambda-1}
\end{equation}
where $i_{1}$ under conditions (\ref{WCA}) is reduced to the form
\begin{equation}
i_{1}=\frac{1}{T^{*}}\int_{0}^{\infty}{\rm d}x\,\frac{x^{2}\sin(x)}{x^{3}+{\kappa^{*}}^{2}\sin(x)},
\label{i-1}
\end{equation}
$\kappa^{*}=\kappa_{D}\sigma$, $\kappa_{D}$ is the  Debye number.

In the PY approximation  $\nu^{{\text HS}}$ is as follows
\[
\nu^{{\text
HS}}=\ln(\eta)+\ln(1-\eta)+\frac{z}{1+z}\ln(z)-\ln(1+z)+\frac{\eta(14-13\eta+5\eta^{2})}{2(1-\eta)^3}.
\]
Taking into account only the first term in
(\ref{deltanu-lambda-1}) we arrive at  the expression for the
chemical potential $\nu_{2}$ in the RPA.  As is seen from
Eq.~(\ref{i-1}), $\Delta\nu_{2}$ does not include the
factor of charge asymmetry explicitly in this case. The second term is the correction to
the RPA resulting from the consideration of the higher-order
correlation effects, namely the third order. It should be noted
that Eq.~(\ref{deltanu-lambda-1}) is obtained in the approximation which is different   from that
considered in \cite{patsahan-mryglod-patsahan:06} (our notation
$\Delta\nu_{2}$ corresponds to $\Delta\nu_{N}$ in
\cite{patsahan-mryglod-patsahan:06}). In particular, the
corresponding formula in
\cite{patsahan-mryglod-patsahan:06} (Eq. (31) in
\cite{patsahan-mryglod-patsahan:06}) includes the third and forth order cumulants.

Putting $z=1$ in (\ref{deltanu-lambda-1})-(\ref{i-1}) we arrive at
the chemical potential of the RPM in the RPA
\begin{equation}
\nu_{2}=\sqrt{2}\left(\nu^{HS}-\frac{1}{2T^{*}}+\frac{i_{1}}{\pi}
\right)
\end{equation}
which reflects the fact of a special symmetry of this model. In
order to go beyond the RPA,   the higher-order correlations  should
be taken into account as it was  done  in
\cite{patsahan_ion}.

Based on  (\ref{deltanu-lambda-1})-(\ref{i-1})  we calculate the
coexistence curves and the corresponding critical parameters for
different values of $z$. The values of the critical parameters
$T_{c}^{*}$ and $\rho_{c}^{*}$ for different $z$ are shown in
Table~2. 
\begin{table}[htbp]
\caption{Critical parameters of the equisize PM with charge
asymmetry for different values of $\lambda$} \vspace{3mm}
\begin{tabular}{ccccc}
\hline \hline
\hspace{7mm}
$z$
\hspace{7mm}
&
\hspace{7mm}
$T_{c}^{*}$
\hspace{7mm}
&
\hspace{7mm}
$10^{2}\rho_{c}^{*}$
\hspace{7mm}
\\
\hline  $1$  & $0.0848$ & $0.907$\\
$2$ &$0.0640$ &$0.720$\\
$3$ &  $0.0469$  &$0.549$ \\
\hline
\hline
\end{tabular}
\end{table}
\begin{figure}[h]
\centering
\includegraphics[height=6.5cm]{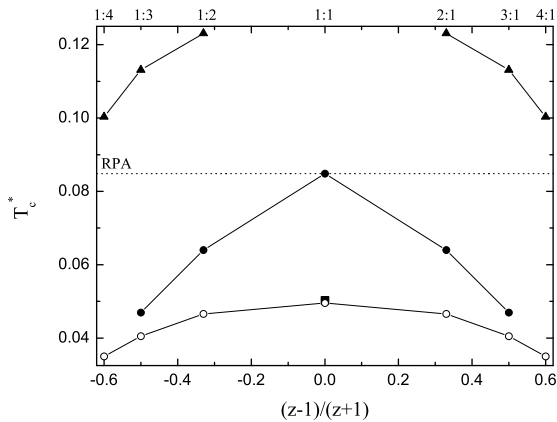}
\caption{Dependence of the critical temperature on charge asymmetry for equisize PMs.
Solid symbols correspond to the results of the CV based theory: circles are the results
based on (\ref{deltanu-lambda-1}-(\ref{i-1}), triangles  are the results from
\cite{patsahan-mryglod-patsahan:06}, $\blacksquare$ is for RPM \cite{patsahan_ion}. Open  circles are the results of simulations: $z=1$ \cite{panagiotopoulos1}, $z=2-3$ \cite{kim-fisher-panagiotopoulos:05}, $z=4$ \cite{Camp-Patay:99}. The dotted line is the result from the RPA.} 
\end{figure}

Figures 3 and 4 show trends of  $T_{c}^{*}(z)$ and $\rho_{c}^{*}(z)$
obtained from (\ref{deltanu-lambda-1})-(\ref{i-1}), together with 
simulation data \cite{Camp-Patay:99,kim-fisher-panagiotopoulos:05}.
The results of \cite{patsahan-mryglod-patsahan:06} are added for
comparison. It should be noted that  the Carnahan-Starling
approximation for the hard-sphere system was used in
\cite{patsahan-mryglod-patsahan:06}. The critical parameters of the
RPM obtained within the framework of this theory but in the
higher-order approximation are shown by the solid squares
($T_{c}^{*}=0.0503$, $\rho_{c}^{*}=0.042$) \cite{patsahan_ion}.
\begin{figure}[h]
\centering
\includegraphics[height=6.5cm]{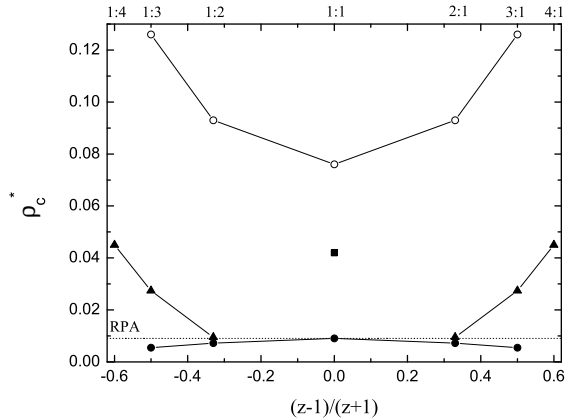}
\caption{Dependence of the critical density
$\rho^{*}=\rho\sigma^{3}$ on  charge asymmetry.  The meaning of the
symbols is the same as in Fig.~4.} 
\end{figure}

As in  \cite{patsahan-mryglod-patsahan:06}, the trend of
$T_{c}^{*}(z)$ obtained  from (\ref{deltanu-lambda-1})-(\ref{i-1})
qualitatively agrees  with simulation findings. Moreover, now the numerical values of
$T_{c}$  are much closer to the simulation data than those found
previously.  On the other hand, the trend of $\rho_{c}^{*}(z)$
found  from (\ref{deltanu-lambda-1})-(\ref{i-1}) is inconsistent with the simulations. It follows from \cite{patsahan-mryglod-patsahan:06} that the  correlation effects of the higher order than the
third order should be taken into account  in order to get the correct trend of the critical density.

\subsection{PMs with size and charge asymmetry}
Now we use formulas (\ref{delta-nu2})-(\ref{nu_s}) for the study
of  the size- and charge-asymmetric PMs with $z=2$ and $\lambda\neq
1$.
As before,  in order to calculate the coexistence curves and the
corresponding critical parameters at different values of $\lambda$
we apply the Maxwell construction. Our results for the critical
parameters $T_{c}^{*}$ and $\rho_{c}^{*}$ are given in Table~3. As
is seen, both the critical temperature and the critical density
decrease with the increase of size asymmetry.

The dependence of the critical parameters on the size asymmetry is
shown graphically in Figs.~5 and~6, respectively, along with the
results of  simulations \cite{yan-pablo:02:2}. In general, the
trends of $T_{c}^{*}$ and $\rho_{c}^{*}$ with $\delta$ are
consistent with the simulation findings: $2$:$1$ systems exhibit a maximum
in both the critical temperature and the critical density when
plotted  as a function of size asymmetry. Similar to simulations,
our results (especially the critical temperature) reveal a pronounced sensitivity to $\delta$. However,
both the critical temperature and the critical density found from
(\ref{delta-nu2})-(\ref{nu_s}) demonstrate maxima at $\delta=0$
($\lambda=1$) while the corresponding maxima obtained  by
simulations are shifted towards nonzero values of $\delta$
($\delta>0$ ). Interestingly, $\rho_{c}^{*}(\delta)$
demonstrates the  general shape similar to that obtained for  the
dumbbell system. As before, our values of the critical density are
more than an order of magnitude lower than those found in the
simulations \cite{yan-pablo:02:2}.
\begin{table}[htbp]
\caption{Critical parameters of the ($2:1$) PM for different
values of $\delta$ ($\delta=(\lambda-1)/(\lambda+1)$)}
\vspace{3mm}
\begin{tabular}{ccc}
\hline \hline\hspace{10mm}  $\delta$\hspace{10mm} &\hspace{10mm}
$T_{c}^{*}$\hspace{10mm} &\hspace{10mm}
$10^{2}\rho_{c}^{*}$\hspace{10mm}
\\
\hline
$-0.67$ & $0.054$& $0.319$\\
$-0.5$  & $0.0587$ & $0.458$
\\
$-0.33$ &$0.0614$ &$0.545$
\\
$-0.2$ &  $0.0630$  &$0.619$
\\
$0$ & $0.0640$& $0.720$
\\
$0.2$& $0.0611$ & $0.616$\\
$0.33$ & $0.0583$ & $0.528$\\
$0.5$ & $0.0553$ & $0.461$\\
$0.67$ & $0.0529$ & $0.436$\\
\hline
\hline
\end{tabular}
\end{table}

\begin{figure}[h]
\centering
\includegraphics[height=6.5cm]{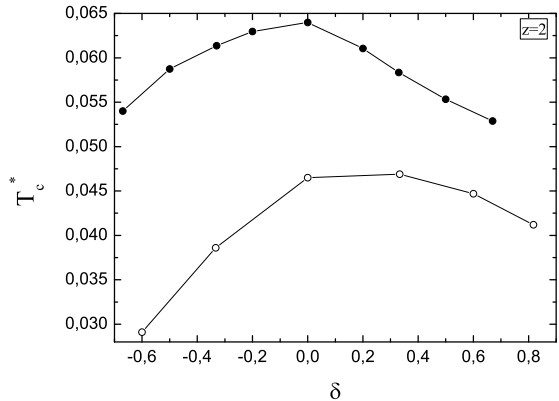}
\caption{Critical temperature of the ($2$:$1$) PM  as a function
of size asymmetry. Solid circles correspond to the results of the CV
based theory; open circles are the results of simulations \cite{yan-pablo:02:2}.} 
\end{figure}
\begin{figure}[h]
\centering
\includegraphics[height=7cm]{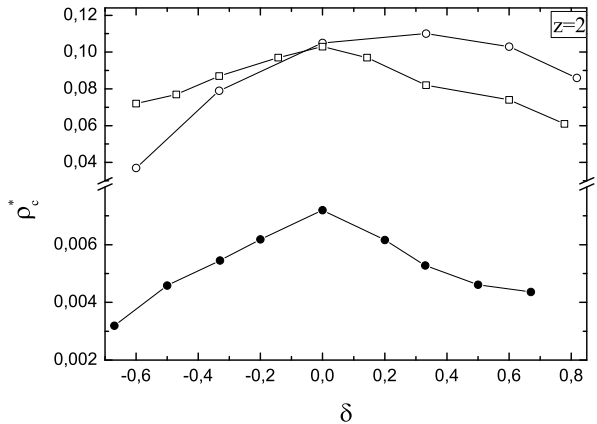}
\caption{Critical density of the ($2$:$1$) PM  as a function of
size asymmetry. Solid circles are the results of the CV based theory.
Open symbols  correspond to the results of  simulations
\cite{yan-pablo:02:2}: circles are  spherical electrolytes;
squares are dumbbell electrolytes.
} 
\end{figure}
\section{Conclusions}
In this paper we have studied the effects of size and charge
asymmetry on the gas-liquid critical parameters of  two-component
PMs using the  CV based theory. The theory  allows one to
take into consideration the effects of  higher-order correlations.
On the other hand,  the
well-known approximations for the free energy, in particular DHLL
and ORPA, can be reproduced within the framework of this theory.
Recently, this approach has been used for the study of the effects of
charge asymmetry on the critical parameters of equisize
charge-asymmetric PMs. It allowed us to calculate, without
additional assumptions (such as the presence of the dipoles or the
higher-order clusters, for example),  the trends of  both the
critical temperature and the critical density with charge
asymmetry that qualitatively agree  with Monte Carlo simulation results
\cite{patsahan-mryglod-patsahan:06}.

First, we have studied the Gaussian approximation of the
functional Hamiltonian of the asymmetric PM. The stability analysis  has led us to the  trends for the critical parameters consistent with those obtained in \cite{Ciach-Gozdz-Stell-07}.  As was shown in \cite{Ciach-Gozdz-Stell-07}, only some of the effects of size and charge asymmetry can be reproduced correctly at this level of consideration.
Then, we study the critical parameters  of size- and
charge-asymmetric PMs  taking into account the higher-order
correlation effects. Following the procedure described in
\cite{patsahan-mryglod-patsahan:06} we have derived in the first
nontrivial approximation an explicit
expression for the  chemical potential conjugate to the
order parameter.  It includes the third-order correlation functions  of the reference system. Then, this expression was used to  study  the three
versions of PM: an equisize PM with charge asymmetry; a monovalent
PM with size asymmetry; a size- and ($2$:$1$) charge-asymmetric
PM.

In conclusion, within the framework of the same approximation we have obtained   the trends of  the
critical temperature and the critical density with  size asymmetry
that qualitatively agree with the Monte Carlo simulation findings: both
$T^{*}_{c}$ and $\rho^{*}_{c}$ decrease with increasing size
asymmetry at the fixed $z$. As regards  the charge asymmetry, the present approximation for the relevant chemical potential  yields a correct trend of the critical temperature with $z$ and  improves the numerical values of $T_{c}^{*}$ when compared with our previous findings. Unlike the results of   \cite{patsahan-mryglod-patsahan:06}, this approximation leads to the opposite trend of the critical density with charge asymmetry although the variation of $\rho_{c}^{*}$ with $z$ is  small.  We expect that the consideration of the correlation effects of higher order  will  enable us  to  correct the trends of the critical density  with  charge asymmetry as well as to improve the numerical values of the critical parameters. This issue  will be considered elsewhere.

\section{Appendices}
\subsection{Explicit expressions for the coefficients $A(k)$, $B(k)$, $C(k)$ and $D(k)$}

\begin{eqnarray*}
 A(k) &=&\frac{1}{\sqrt{1+\alpha_{1}^{2}}}, \qquad
 B(k) =\frac{\alpha_{1}}{\sqrt{1+\alpha_{1}^{2}}}, \nonumber \\
 C(k) &=&\frac{1}{\sqrt{1+\alpha_{2}^{2}}}, \qquad
 D(k)=\frac{\alpha_{2}}{\sqrt{1+\alpha_{2}^{2}}},
\end{eqnarray*}
where
\begin{eqnarray*}
\alpha_{1,2}=\frac{\tilde{\mathcal C}_{--}(k)-\tilde{\mathcal
C}_{++}(k)\pm \sqrt{(\tilde{\mathcal C}_{++}(k)-\tilde{\mathcal
C}_{--}(k))^{2}+4\tilde{\mathcal C}_{+-}(k)^{2}}}{2\tilde{\mathcal
C}_{+-}(k)}.
\end{eqnarray*}

\subsection{Expressions for ${\mathfrak{M}}_{\alpha_{1}\alpha_{2}}(0)$ and
${\mathfrak{M}}_{\alpha_{1}\alpha_{2}\alpha_{3}}(0,0)$ in the PY
approximation}

Using the Lebowitz' solution of the PY equation \cite{leb1} for a
hard sphere system one can obtain the explicit expressions for the
Fourier transforms of Ornstein-Zernike (OZ) direct correlation
functions $\tilde c_{\alpha\beta}^{HS}(k)$. In the
long-wavelength limit they are of the form \cite{patyuk4}:
\begin{eqnarray}
\rho_{+}\tilde c_{++}^{HS}(0)& =& - 2\eta_{+}(4a_{+} + 3\beta_{+}
+ 2\gamma_{+}), \label{c11.a}
\\
\rho_{-}\tilde c_{--}^{HS}(0)&  = & - 2\eta_{-}(4a_{-} +
3\beta_{-} + 2\gamma_{+}\lambda^{-3}), \label{c22.b}
\\
\sqrt{\rho_{+}\rho_{-}}\tilde c_{+-}^{HS}(0) & = & -\frac{1}{5!}\{
A + B[ 10
\beta_{+-}(4\tilde a + 3) + \nonumber \\
&& + 6\gamma_{+-}(5\tilde a + 4) + 4\gamma_{+}(6\tilde a + 5)]\},
\label{c12.c}
\end{eqnarray}
where the following notations are introduced
\begin{eqnarray}
\eta = \eta_{+} + \eta_{-}, \quad
\eta_{+}&=&\frac{x_{+}\eta\lambda^{3}}{x_{-}+x_{+}\lambda^{3}},
\qquad \eta_{-}=\frac{x_{-}\eta}{x_{-}+x_{+}\lambda^{3}},
\label{b3.3a}
\end{eqnarray}
\begin{eqnarray}
\tilde a = \frac{1 - \lambda}{2\lambda},\quad
h=\eta\frac{\sqrt{x_{+}x_{-}}}{x_{-}+x_{+}\lambda^{3}}, \quad
x_{i}=\frac{\rho_{i}}{\rho},
 \label{eta-i}
\end{eqnarray}
\begin{eqnarray}
a_{+}&  =&  \frac{1}{(1 - \eta)^{4}}\{1 - \eta^{3} + (\eta_{+} +
\lambda^{3}\eta_{-})(\eta^{2} + 4(1 + \eta)) \nonumber \\
&&  -3 \eta_{-}(1 -\lambda)^{2}[(1 + \eta_{+} + \lambda(1 +
\eta_{-}))
(1 - \eta + 3\eta_{+}) \nonumber \\
&& + \eta_{+}(1 - \eta)]\},
\end{eqnarray}
\begin{eqnarray}
a_{-} & =&  \frac{1}{\lambda^{3}(1 - \eta)^{4}}\{\lambda^{3}(1 -
\eta^{3}) +
(\eta_{+} + \lambda^{3}\eta_{-})(\eta^{2} + 4(1 + \eta)) \nonumber \\
&& - 3\eta_{+}(1 -\lambda)^{2}[(1 + \eta_{+} + \lambda(1 +
\eta_{-}))
(1 - \eta + 3\eta_{-})  \nonumber \\
&& + \lambda\eta_{-}(1 - \eta)]\}, \label{b3.3b}
 \end{eqnarray}
 \begin{eqnarray}
\beta_{+}&  = & -6\left[\eta_{+} g_{++}^{2} + \frac{1}{4}\eta_{-}(1 +
\lambda)^{2}
\lambda g_{+-}^{2}\right],   \nonumber \\
\beta_{-} & =&  -6\left[\eta_{-} g_{--}^{2} +
\frac{1}{4}\eta_{+}\lambda^{-3}
(1 + \lambda)^{2} g_{+-}^{2}\right],  \nonumber \\
\beta_{+-} & = & -3\lambda(1 - \lambda)(\lambda^{-2}\eta_{+}
g_{++} + \eta_{-} g_{--}) g_{+-}, \label{b3.3c}
\end{eqnarray}
\begin{eqnarray}
\gamma_{+} & =&  \frac{1}{2}\left(\eta_{+} a_{+} + \lambda^{3}\eta_{-}
a_{-}\right),
\nonumber  \\
\gamma_{-} & =&\frac{\gamma_{+}}{\lambda^{3}},     \qquad
\gamma_{+-} = 2\gamma_{+}\frac{1 - \lambda}{\lambda},   \label{b3.3d}  \\
g_{++} & =&  \frac{1}{(1-\eta)^{2}}\left[1 + \frac{\eta}{2} +
\frac{3}{2}\eta_{-}
(\lambda - 1)\right],  \nonumber \\
g_{--} & = & \frac{1}{(1-\eta)^{2}}\left[1 + \frac{\eta}{2} +
\frac{3}{2}\eta_{+}
(\lambda^{-1} - 1)\right],  \nonumber \\
g_{+-} & =&  \frac{1}{(1-\eta)^{2}}\left[1 + \frac{3\eta(1 - \lambda)}
{4(1 + \lambda)}(\eta_{+} - \eta_{-})\right]. \label{b3.3e}
\end{eqnarray}
\begin{equation}
A = \frac{5a_{+}(1 + \lambda)^{3}B}{\lambda^{3}}, \qquad B =
4!\sqrt{\lambda^{3}\eta_{+}\eta_{-}}.   \label{b3.3f}
\end{equation}
The expressions for
$S_{\alpha\beta}(k)={\mathfrak{M}}_{\alpha\beta}(k)/\sqrt{N_{\alpha}N_{\beta}}$
can be found from the OZ equations
\begin{equation}
S_{++}(k) =  \frac{1 - \rho_{-}\tilde c_{--}^{HS}(k)}{(1 -
\rho_{+} \tilde c_{++}^{HS}(k))(1 - \rho_{-}\tilde c_{--}^{HS}(k))
- \rho_{+}\rho_{-} (\tilde c^{HS}_{+-}(k))^{2}}, \label{b3.2a}
\end{equation}
\begin{equation}
S_{--}(k) =  \frac{1 - \rho_{+}\tilde c_{++}^{HS}(k)}{(1 -
\rho_{+} \tilde c_{++}^{HS}(k))(1 - \rho_{-}\tilde c_{--}^{HS}(k))
- \rho_{+}\rho_{-} (\tilde c_{+-}^{HS}(k))^{2}}, \label{b3.2b}
\end{equation}
\begin{equation}
S_{+-}(k) = \frac{\sqrt{\rho_{+}\rho_{-}}\tilde c_{+-}^{HS}(k)}{(1
- \rho_{+} \tilde c_{++}^{HS}(k))(1 - \rho_{-}\tilde
c_{--}^{HS}(k)) - \rho_{+}\rho_{-}(\tilde c_{+-}^{HS}(k))^{2}}.
 \label{b3.2c}
\end{equation}
Eqs (\ref{c12.c})-(\ref{b3.2c}) should be supplemented by the
electroneutrality condition.

The explicit expressions for
$S_{+++}(0,0)={\mathfrak{M}}_{+++}(0,0)/N_{+}$ and
$S_{++-}(0,0)={\mathfrak{M}}_{++-}(0,0)/N_{+}$ can be obtained
from the relations
\begin{eqnarray}
S_{+++}(0,0) & = & S_{++}(0)\left[S_{++}(0) +\eta_{+}
\left(\frac{\partial S_{++}(0)}{\partial
\eta_{+}}\right)_{\eta_{-}}\right] \nonumber
\\& &
+  \eta_{-}\sqrt{\frac{x_{+}}{x_{-}}}S_{+-}(0)
\left(\frac{\partial S_{++}(0)}{\partial
\eta_{-}}\right)_{\eta_{+}}, \label{S111}
\end{eqnarray}
\begin{eqnarray}
S_{++-}(0,0) & = & \sqrt{\frac{x_{-}}{x_{+}}}\left[S_{++}(0)
+\eta_{+} \left(\frac{\partial S_{++}(0)}{\partial
\eta_{+}}\right)_{\eta_{-}}\right] \nonumber
\\& &
+ \eta_{-}S_{--}(0)\left(\frac{\partial S_{++}(0)}{\partial
\eta_{-}}\right)_{\eta_{+}}. \label{S112}
\end{eqnarray}
The expressions for ${\mathfrak{M}}_{---}(0,0)$ and
${\mathfrak{M}}_{+--}(0,0)$ can be obtained replacing indices
``$+$'' by indices ``$-$'' and vice verse.

Final formulas should be supplemented by the electroneutrality
condition.

\subsection{Explicit expressions for $\nu_{2}^{S}$ and $\nu_{2}^{\text
 HS}$}
We obtain for $\nu_{2}^{S}$
\begin{equation*}
\nu_{2}^{S}=\frac{(1+\lambda)(z+\lambda)}{4T^{*}\lambda\sqrt{1+z^{2}}}.
\end{equation*}
In the PY approximation  $\nu_{2}^{\text HS}$ has the form 
\begin{eqnarray*}
 &&\nu_{2}^{\text
 HS}=\frac{1+z}{\sqrt{1+z^{2}}}\left[\ln\eta+\ln(1-\eta)
 +\frac{z}{1+z}\ln(z)-\ln(z+\lambda^{3})
 +\frac{\eta(1+\eta+\eta^{2})}{(1-\eta)^{3}}\right.
 \\
 &&
 \left.-\frac{3\eta}{2(1+z)(1-\eta)^{3}(z+\lambda^{3})^{2}}\left[2z\eta(1-\lambda)^{2}\left((1+\lambda)(z+\lambda^{3})+
 \eta\lambda(z+\lambda^{2})\right)\right.\right.\\
&& \left.\left.
-4(z+\lambda)(z+\lambda^{2})(z+\lambda^{3})(1-\eta)^{2}-3\eta(z+\lambda^{2})^{3}(1-\eta)
\right]
\right].
\end{eqnarray*}

\end{document}